\documentclass{emulateapj}

\newcommand{\gsim}{\hspace{0.3em}\raisebox{0.4ex}{$>$}\hspace{-0.75em}\raisebox{-.7ex}{$\sim$}\hspace{0.3em}}
\newcommand{\lsim}{\hspace{0.3em}\raisebox{0.4ex}{$<$}\hspace{-0.75em}\raisebox{-.7ex}{$\sim$}\hspace{0.3em}}

\shorttitle{A Subaru/Suprime-Cam Survey of NGC55's Stellar Halo}
\shortauthors{Tanaka et al.}

\begin{document}


\title{STRUCTURE AND POPULATION OF THE NGC55 STELLAR HALO FROM A
SUBARU/SUPRIME-CAM SURVEY\altaffilmark{1}} 


\author{Mikito~Tanaka\altaffilmark{2},
        Masashi~Chiba\altaffilmark{2},
        Yutaka~Komiyama\altaffilmark{3},
        Puragra~Guhathakurta\altaffilmark{4}, 
	and 
	Jason~S.~Kalirai\altaffilmark{5}}


\altaffiltext{1}{Based on data collected at the Subaru Telescope, which
 is operated by the National Astronomical Observatory of Japan.}
\altaffiltext{2}{Astronomical Institute, Tohoku University, Aoba-ku,
 Sendai 980-8578, Japan (current address); mikito@astr.tohoku.ac.jp} 
\altaffiltext{3}{National Astronomical Observatory of Japan, 2-21-1
 Osawa, Mitaka, Tokyo 181-8588, Japan}
\altaffiltext{4}{University of California Observatories/Lick
 Observatory, University of California Santa Cruz, 1156 High Street,
 Santa Cruz, California 95064, USA} 
\altaffiltext{5}{Space Telescope Science Institute, Baltimore, MD 21218} 


\begin{abstract}
As part of our survey of galactic stellar halos, we investigate the
structure and stellar populations of the northern outer part of the stellar halo
in NGC~55, a member galaxy of the Sculptor Group, 
using deep and wide-field $V$- and $I$-band images taken with
Subaru/Suprime-Cam. Based on  the analysis of the color-magnitude diagrams (CMDs)
for red-giant-branch (RGB) stars,
we derive a tip of RGB (TRGB)-based distance
modulus to the galaxy of $(m-M)_0 = 26.58 \pm 0.11 (d = 2.1 \pm 0.1 {\rm Mpc})$. 
From the stellar density maps, 
we detect the asymmetrically disturbed, thick disk structure and two metal-poor 
overdense substructures in the north region of NGC~55, which may correspond to 
merger remnants associated with hierarchical formation of NGC~55's halo. 
In addition, we identify a diffuse metal-poor halo extended out to at least $z \sim 16$~kpc
from the galactic plane.
The surface-brightness profiles toward the $z$-direction perpendicular to 
the galactic plane suggest that the stellar density distribution 
in the northern outer part of NGC~55 is described by a locally isothermal 
disk at $z \lsim 6$~kpc and a likely diffuse metal-poor halo
with $V$-band surface brightness of $\mu_{\rm V} \gsim 32$ mag arcsec$^{-2}$,
where old RGB stars dominate.
We derive the metallicity distributions (MDs) of these structures 
on the basis of the photometric comparison of RGB stars with the theoretical 
stellar evolutionary models. The MDs of the thick disk structures show 
the peak and mean metallicity of [Fe/H]$_{\rm peak} \sim -1.4$ and
[Fe/H]$_{\rm mean} \sim -1.7$, respectively,
while the outer substructures show more metal-poor features than
the thick disk structure.
Combined with the current results with our previous study for M31's halo,
we discuss the possible difference in the formation process of stellar halos
among different Hubble types.

\end{abstract}





%%% Section 1 %%%

\section{Introduction}\label{sec:intro}

Our understanding of how a stellar halo surrounding a disk galaxy like
the Milky Way has formed is still enigmatic, because the observational
information of this faint galactic component is yet limited, except for
Local Group galaxies. Halo stars in the Milky Way are characterized by
low metal abundance and high velocity dispersion and this extreme nature
reflects the early chemo-dynamical evolution of the Milky Way. Extensive
analyses of these stars have revealed, e.g., the dual nature of the halo
structure (inner flattened halo in prograde rotation and outer spherical
halo in retrograde rotation) \citep[e.g.,][]{Carollo07}. Also, recent growing
observational evidence suggests that the Milky Way halo has formed from
an assembly process of many subsystems, as deduced from stream-like halo
substructures \citep[e.g.,][]{Yanny00}. Similar substructures have also been
identified in M31's halo, including the Giant Stellar Stream
\citep{Ibata07}. Indeed, M31's halo provides a clear external view of
ancient stellar populations: the halo is found to be extended more than
150~kpc \citep{Raja05} and has characteristic spatial structure and
metallicity distributions \citep[e.g.,][]{Kalirai06,Ibata07,Tanaka10}.
Another Local Group galaxy,
M33, seems to have a rather smooth halo, although its disk orientation
prevents the more refined picture of the halo. Thus, detailed studies of
old stellar halos in various-type galaxies provide us with important
clues to the understanding of galaxy formation. 

Exploring stellar halos even beyond the Local Group is of great
importance to accomplish our understanding of their generic nature and
past formation history in different disk galaxies at different
environments. Model investigations \citep[e.g.,][]{Kauffmann96} predict that each
disk galaxy has been developed through a different formation and
evolutionary path: the collapse epoch, star formation history, and
assembly rate of subsystems are not common and each stellar halo 
is expected to hold a different morphology, as other components (bulge
and disk) differ along the Hubble sequence. 

In this regard, an edge-on spiral SB(s)m galaxy, NGC~55, which is a member of the nearby 
Sculptor group that consists of approximately 30 galaxies \citep{Cote97,Jerjen00}, 
is an excellent test-bed for this study: it provides an external perspective of a nearby late-type 
disk galaxy and yet is close enough to resolve individual stars. 
For example, NGC~55 shows asymmetric extra-planar morphology as derived from emission-line images
\citep{Ferguson96} and its nuclear source shows, based on far-infrared measurements,
energetic star formation rate, which is of comparable luminosity to
the bright star formation regions at the center of M33 and the Milky Way \citep{Engelbracht04}.
Based on the observational study of asymptotic giant branch (AGB) stars,
\citet{Davidge05} found that there are no evidence for a young or intermediate-age component
in the extraplanar region such as 2~kpc off of the disk plane,
suggesting that the population has ages of about 10~Gyr.
On the other hand, \citet{Tikhonov05} found an exponential spatial distribution of RGB stars
between 2 and 7~kpc from the center along the minor axis, using HST/WFPC2. 
\citet{Seth05a,Seth05b}, as part of a snapshot survey of 16 nearby edge-on late-type 
galaxies, observed the thin disk of NGC~55 using HST/ACS. They determined the 
fundamental properties of the resolved stellar population such as the distance, 
MD and spatial distributions of main-sequence, AGB and RGB stars 
in the inner part at $z\lsim1.5$~kpc. However, since all the previous studies using HST 
are restricted to a small field-of-view, the global structure of an outer part of NGC~55 is 
yet unknown. Therefore, we conduct a high-resolution, 
deep wide-field survey of the halo of NGC~55 using Subaru/Suprime-Cam, 
to elucidate its fundamental properties for the first time. 

The layout of this paper is as follows. In Section~\ref{sec:data}, we
present observations and summarize procedures for performing photometry
for our Suprime-Cam data. Sections~\ref{sec:dist}, \ref{sec:map} and
\ref{sec:metal} are devoted to our results for quantitative analysis of
the CMDs of NGC~55's stellar populations, including the distance
estimate using the TRGB stars, the detection of
substructures in NGC~55's halo from the stellar population maps,
the radial profiles of the resolved stars, and
their MDs using the theoretical isochrones. In Section~\ref{sec:dis}, 
comparing with previous studies on stellar halos, we
discuss the implications of our results and conclusions are drawn.

%%% Section 2 %%%
\section{Data}\label{sec:data}

\begin{figure}[htpd]
 \epsscale{1}
 \plotone{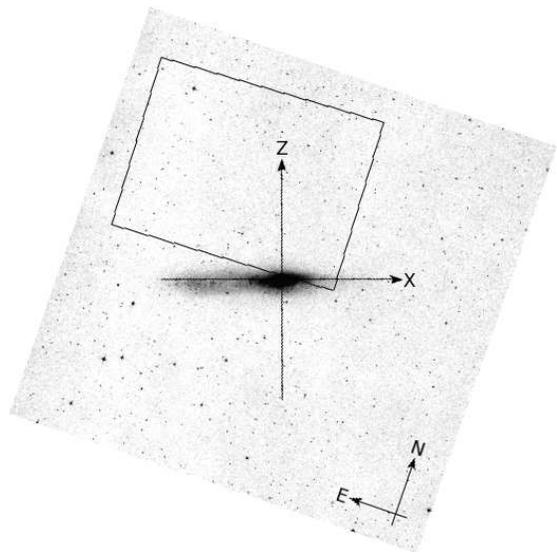}
 \caption[a]{
 The locations and field of view of our Subaru/Suprime-Cam field,
 overlaid on a red Digitized Sky Survey image of NGC~55 covering about
 $60\arcmin \times 60\arcmin$. A rectangle shows an area in which we
 analyzed our Suprime-Cam data, corresponding to 0.22 square degrees.
 }
 \label{fig:map}
\end{figure}

In this study, we use the Suprime-Cam imager \citep{Miyazaki02} on the
8.2-m Subaru Telescope on Mauna Kea in Hawaii. Suprime-Cam consists of
ten $2048\times4096$ CCDs with a scale of $0\farcs202$ per pixel and
covers a total field-of-view of $34\arcmin \times 27\arcmin$. 
We have observed the north part of a stellar halo in NGC~55
in 2009 December. Figure~\ref{fig:map} shows the location of
our Suprime-Cam pointing containing the edge of NGC~55's thick disk
suggested by \citet{Tikhonov05} and the outer halo out to around 17~kpc
perpendicular to the galactic plane. In this study, we define that
$x$-axis corresponds to the galactic major axis, and $z$-axis which is
converted based on the inclination ($80\arcdeg$) in a direction
perpendicular to the galactic mid-plane. The observations were made with
Johnson $V$-band and Cousins $I$-band filters. Exposure times of our
targeted fields are 960 sec and 1800 sec in $V$ and $I$-band,
respectively. The weather condition was photometric but slightly poor, 
with seeing of around $1\farcs0$. 
To estimate the number of foreground stars of the Milky Way 
and unresolved background galaxies in each NGC~55's field, 
we also obtained the imaging data for the control field, which is located at the 
same galactic latitude ($b \sim -76\arcdeg$) as the object field, but is 
about $3\arcdeg$ away from it.

We have performed reduction, photometry and artificial star experiments
following the same manner as given in \citet{Tanaka10}. The raw data were reduced
in the standard procedures, with the software package SDFRED, a useful
pipeline developed to optimally deal with Suprime-Cam images
\citep{Yagi02,Ouchi04}. We then conducted PSF-fitting photometry using
the IRAF version of the DAOPHOT-I\hspace{-.1em}I software
\citep{Stetson87}. Comparing input artificial stars with output ones, we
evaluated incompleteness due to low S/N ratio and selection criteria
based on DAOPHOT parameters. Then, the 50\% (80\%) completeness of 
the object field for each filter is reached at $V_{\rm NGC55}^{50}=25.61$ 
($V_{\rm NGC55}^{80} = 25.03$) and $I_{\rm NGC55}^{50}=25.07$ 
($I_{\rm NGC55}^{80}=24.40$) 
in the nearest field to the NGC~55 center, while the completeness of the control field 
reached at $V_{\rm Control}^{50}=25.80$ ($V_{\rm Control}^{80} = 25.23$) 
and $I_{\rm Control}^{50}=25.27$ ($I_{\rm Control}^{80}=24.61$) .
The typical mean magnitude errors at
the 50\% (80\%) completeness limits, based on the simulated numerous
stars, are of the order of $\sigma_V \sim 0.3$ (0.2) and $\sigma_I \sim
0.3$ (0.2). 

Brief information about NGC~55 and the references are given in
Table~\ref{tab:n55}. Reddening correction is applied to this Suprime-Cam
field based on the extinction maps of \citet*{Schlegel98}, and the
\citet{Dean78} reddening law $E(V-I)=1.34E(B-V)$ and $A_I=1.31E(V-I)$. 
The central coordinate of NGC~55 is derived from 2MASS observations,
in agreement with the peak position of the blue light and also the
location of maximum symmetry of H$_{\rm I}$ velocity field as described
in \citet{Hummel86}. In addition, the galaxy's circular velocity, inclination and
position angle are derived from their neutral hydrogen and radio
continuum observations.

\begin{deluxetable}{lcc}
 \tablewidth{0pt}
 \tabletypesize{\footnotesize}
 \tablecaption{Properties of NGC~55 \label{tab:n55}}
 \tablehead{
 \colhead{Property} & \colhead{Value} & \colhead{Reference}
 }
 \startdata
 RA (J2000)         & $00^{\rm h}14^{\rm m}53\fs6$ & NED \\
 Dec (J2000)        & $-39{\arcdeg}11{\arcmin}48{\farcs}0$ & NED \\
 Morphological type & SB(s)m:sp & NED \\
 A$_I$              & 0.024 & \citet{Schlegel98} \\
 $V_{\rm circ}$~(km\,s$^{-1}$)     & 110 & \citet{Hummel86} \\
 Inclination        & 80{\arcdeg} & \citet{Hummel86} \\
 Position Angle     & 109{\degr} & \citet{Hummel86} \\
 $(m-M)_0$          & 26.58 (2.1~Mpc) & this work \\
% ${\rm [Fe/H]}_{\rm med}$ & $-1.5$ & this work \\
 \enddata
\end{deluxetable}

%%% Section 3 %%%

\section{Distance to NGC~55}\label{sec:dist}

\begin{figure*}[htpd]
 \epsscale{1}
 \plotone{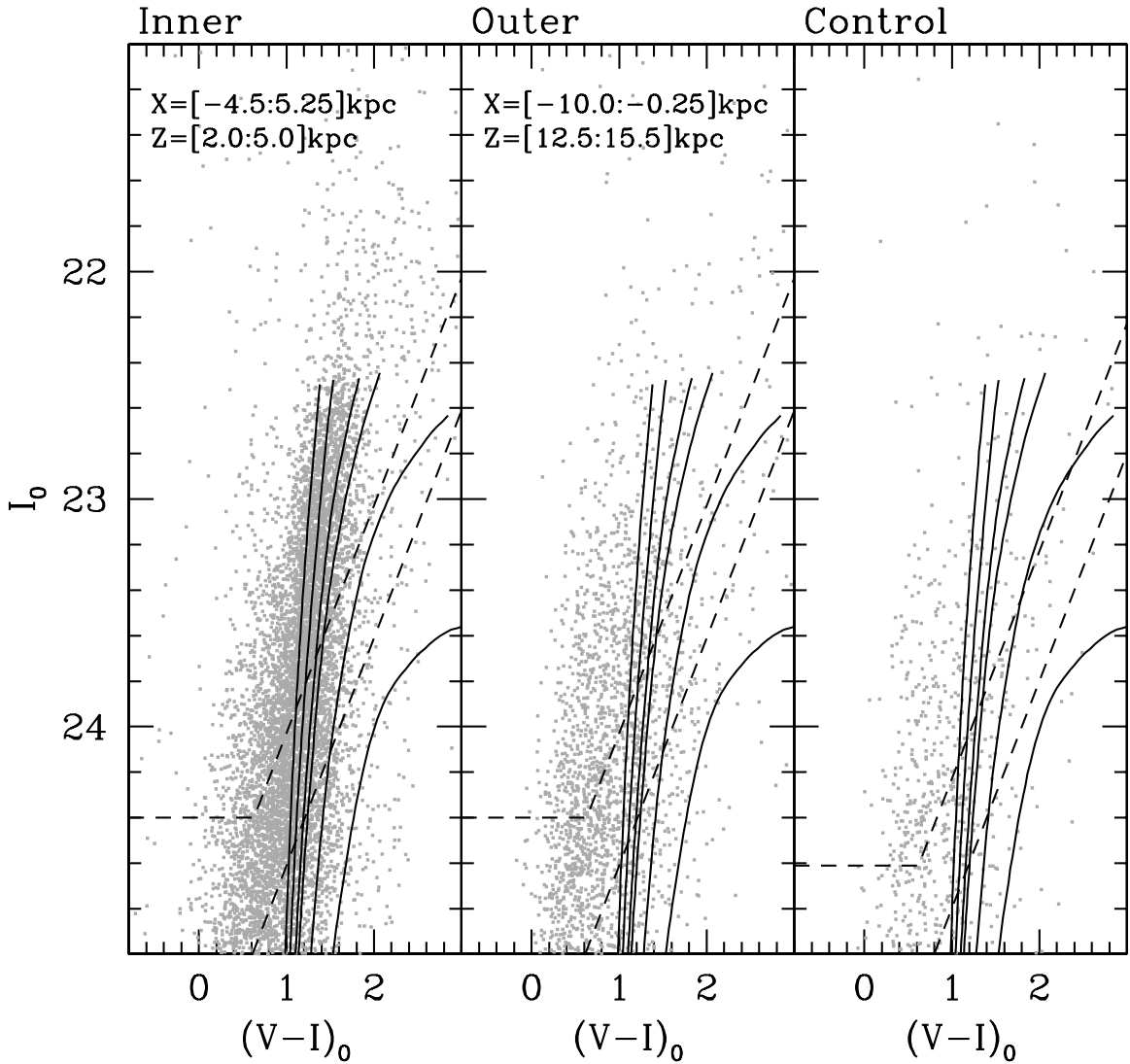}
 \caption[a]{[$(V-I)_0$, $I_0$] color-magnitude diagrams for stellar-like
 sources in the inner and outer regions of NGC~55's halo (left and
 middle) and the control field (right). The solid lines are theoretical
 isochrones \citep{VandenBerg06} of age 12 Gyr and
 [$\alpha$/Fe]$=+0.3$ spanning the metallicity range [Fe/H]$ = -2.31$,
 $-1.71$, $-1.31$, $-1.14$, $-0.71$ and $-0.20$. The dashed lines denote
 the full ranges of the 50\% and 80\% completeness levels.}
 \label{fig:cmd}
\end{figure*}

Figure~\ref{fig:cmd} shows the typical CMDs for stellar-like sources in
the inner and outer region of NGC~55's halo (left and middle) and for
the control field (right), after removing extended sources such as
background galaxies and cosmic rays based on DAOPHOT parameters. The
three CMDs have the same sky coverage of $0.022$ square degrees. The
solid lines in the CMDs show theoretical RGB tracks from
\citet{VandenBerg06} for an age of 12 Gyr, [$\alpha$/Fe]$ = +0.3$, and
metallicities (left to right) of [Fe/H]$ = -2.31, -1.71, -1.31, -1.14,
-0.71$ and $-0.30$. The dashed lines denote about 50\% and 80\%
completeness limits as determined by artificial star experiments.

The clearest feature in the inner region of
NGC~55's halo is a metal-poor RGB with $-2.3 < {\rm [Fe/H]} < -1.1$. 
Furthermore, at $I \sim 22.5$, we also recognize a tip of RGB (TRGB). 
The stars distributed in the brighter part than the TRGB magnitude are probably
metal-rich and/or young stellar population in the thermally
pulsating AGB phase (TPAGB). 
In contrast, it seems that the CMD of the outer region is rather similar to
that of the control field, suggesting that there might be only a little
intrinsic halo population in this outer field beyond $z \sim 12$ kpc
from the NGC~55 center. 

The TRGB is a useful indicator to estimate the distance to resolved
galaxies \citep{Salaris05}. If there are a sufficient number of old and
metal-poor RGB stars in a targeted field, the TRGB is easily detected as
a sharp cutoff of the luminosity function (LF) with the application of
an edge-detection algorithm, the Sobel filter
\citep[e.g.,][]{Salaris05,Tanaka10}. 

\begin{figure}[htpd]
 \begin{center}
  \epsscale{1}
  \plotone{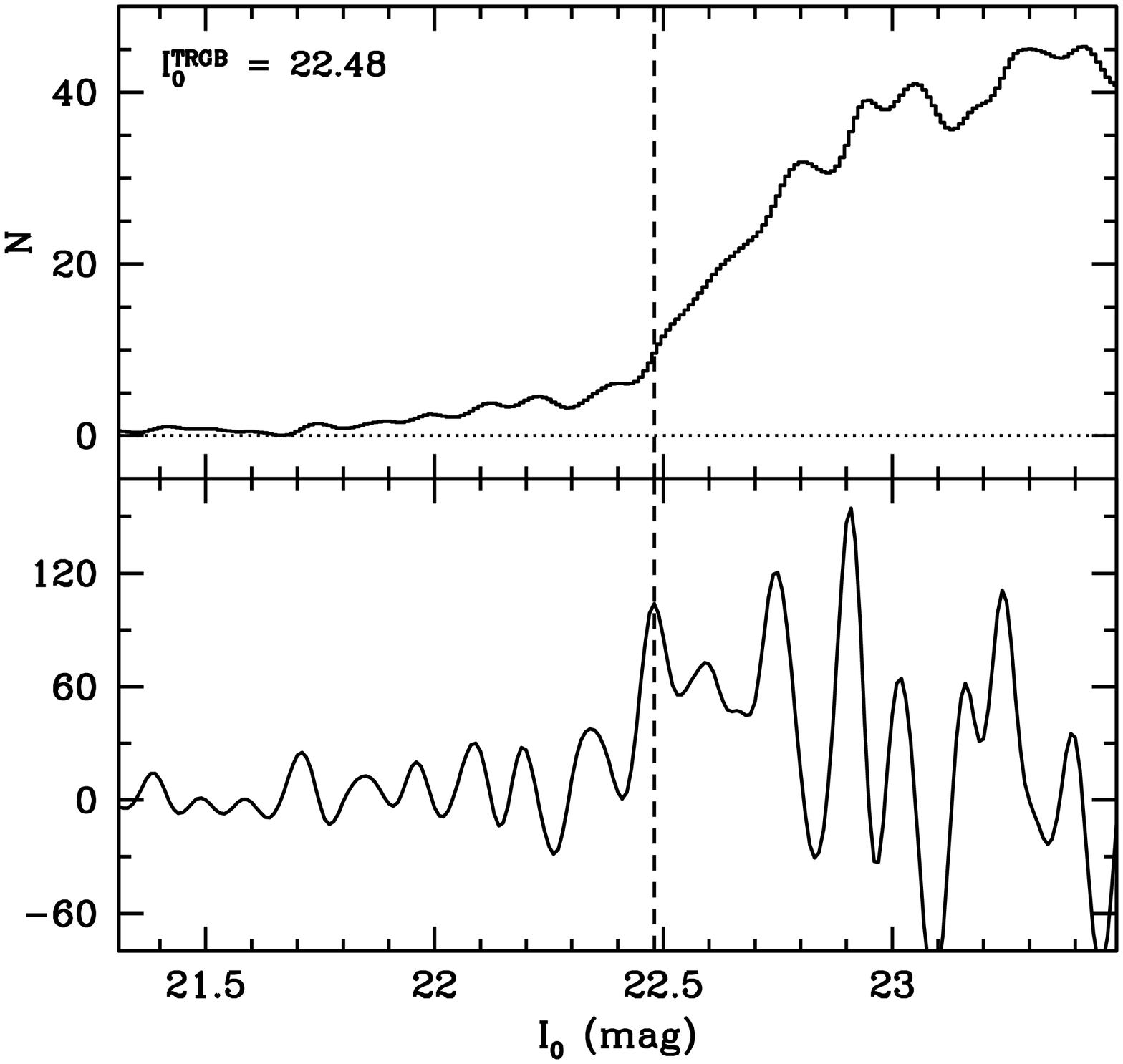}
  \caption[a]{TRGB detection in NGC~55. 
  The upper panel presents the smoothed LFs as a function of the
  incompleteness-corrected and background-subtracted $I_0$-band
  magnitude zoomed in around the TRGB magnitude. The lower panel
  indicates the Sobel filter response to the LF. The vertical dashed
  line shows the derived TRGB magnitude.
  }
  \label{fig:trgb}
 \end{center}
\end{figure}

Detection of the TRGB is shown in Figure~\ref{fig:trgb}. The LF around
the TRGB magnitude is derived from the selected RGB stars in a lower
metallicity range than [Fe/H]$ = -0.7$ to minimize the effect of
empirical calibration of the TRGB magnitude. The LF is gradually standing
up at $I_0 \sim 21.7$ because of the increase of TPAGB stars and steeply
rising up at $I_0 = 22.48 \pm 0.05$. 
The TRGB magnitude suffers from statistical uncertainty of the order of
about 0.05 mag. In estimating the extinction-corrected apparent magnitude of TRGB,
systematic errors also arise in association with zero point uncertainties,
aperture corrections, photometric errors, smoothing of the LFs and extinction law;
the resultant total error is evaluated by an rms of these errors.
We further conduct 1,000 Monte Carlo realizations of the data,
such that the brightness of each object is re-distributed
in the form of a Gaussian distribution with standard deviation equal to
the photometric error. It is found that the resultant deviation of the TRGB magnitude
is only as small as 0.03 mag and thus reasonably small.

On the assumption that the absolute $I_0$-band magnitude of the TRGB is
$M_I^{\rm TRGB} = -4.1 \pm 0.1$ for metal-poor TRGB stars, the distance
modulus \citep[e.g.,][]{Salaris05} is $26.58 \pm 0.11$ corresponding to
$2.1 \pm 0.1$~Mpc. This value is consistent with previous studies based
on the same TRGB method as this study \citep{Seth05a,Tikhonov05}. 

%%% Section 4 %%%

\section{Stellar Population Maps}\label{sec:map}

In this section, we investigate how the stellar populations of
NGC~55's halo are distributed spatially out to about 17 kpc from 
NGC~55's center. To do so, we draw stellar density maps corresponding to
each stellar population, based on the matched filter method explained in
our previous study \citep{Tanaka10}. First, we visually separate the two
major structures (thick disk and diffuse halo) and some substructures
from the following spatial distributions of stellar population of
NGC~55. Then, we quantitatively discriminate the difference of their
populations based on their CMDs and MDs.

\begin{figure*}[htpd]
 \begin{center}
  \epsscale{1}
  \plottwo{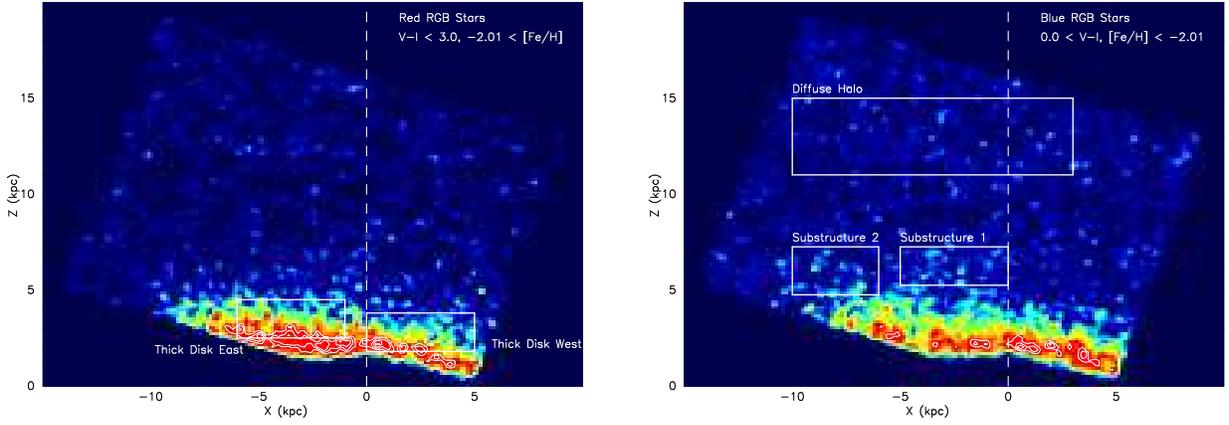}{./f4b.ps}
  \caption[a]{Log-scaled matched filter maps to a limiting magnitude of
  $I_0 = 24.4$ and $V_0 = 25.0$. The left panel shows the distribution of  
  red RGB stars chosen with $(V-I) < 3.0$ and $-2.01 < $[Fe/H], while the
  right panel is for blue RGB stars with
  $0.0<(V-I)$ and [Fe/H]$ < -2.01$. The resolution of this map is
  $0.15\degr \times 0.15\degr$ pixels, smoothed with a Gaussian kernel
  over 3 pixels.
  }
  \label{fig:map_rgb}
 \end{center}
\end{figure*}

Figure~\ref{fig:map_rgb} shows 
density maps of RGB stars in the north halo region of NGC~55, divided
[Fe/H] into two non-overlapping ranges (left panel: [Fe/H]$ < -2.01$,
right panel: $-2.01 < $[Fe/H], see also Figure~\ref{fig:cmd}). 
Then, we impose several criteria to get rid
of unpleasant contaminations unrelated to the RGB stars from our
photometric catalog: for example, appropriate color, magnitudes
between the TRGB and 80\% completeness. In addition, to remove the
remaining background and foreground contaminations, we subtract the
constant value estimated based on the data of the control field from
each bin of the stellar maps. The background levels for RGB and AGB stars
are $\mu_{V, \rm back}^{\rm RGB} \sim 30.8 \pm 0.1$ mag arcsec$^{-2}$ and 
$\mu_{V, \rm back}^{\rm AGB} \sim 33.6 \pm 0.1$ mag arcsec$^{-2}$, respectively.

\subsection{Inner Structures}\label{sec:inner}

Both panels of Figure~\ref{fig:map_rgb} show that NGC~55 has an asymmetric stellar structure in
eastern and western regions, regardless of metal abundances. There is
a somewhat extended structure towards the $z$-direction in the eastern
part. We nominally refer the eastern and western structures to as ``Thick Disk East (TDE)''
and ``Thick Disk West (TDW)'', respectively hereafter. TDE seems to be 
related to the disturbed eastern disk which is in the shape of a
tadpole tail, as seen in Figure~\ref{fig:map}. This might suggest
that an accreted satellite galaxy gave a structural influence
on the eastern part of the disk structure \citep[e.g.,][]{Abadi03,Yoachim06}.

In contrast, assuming that TDW, which is a high density region in the
north-western part of NGC~55, is undisturbed, we regard this
region as the normal thick disk. The scale height of the plausible thick
disk of NGC~55 is estimated as $z_{0,thick} \sim 1.6$~kpc (see also
Figure~\ref{fig:sbz}), which is consistent with that
estimated by the previous HST study \citep{Tikhonov05}. In fact,
considering $V_{\rm circ} = 110$~km~s$^{-1}$, the estimated scale height is in good
agreement with the relation between galactic rotational velocity and
scale-height of thin/thick disk constructed by \citet{Yoachim06} (see
their Fig.~9). 

\begin{figure*}[htpd]
 \begin{center}
  \epsscale{1}
  \plottwo{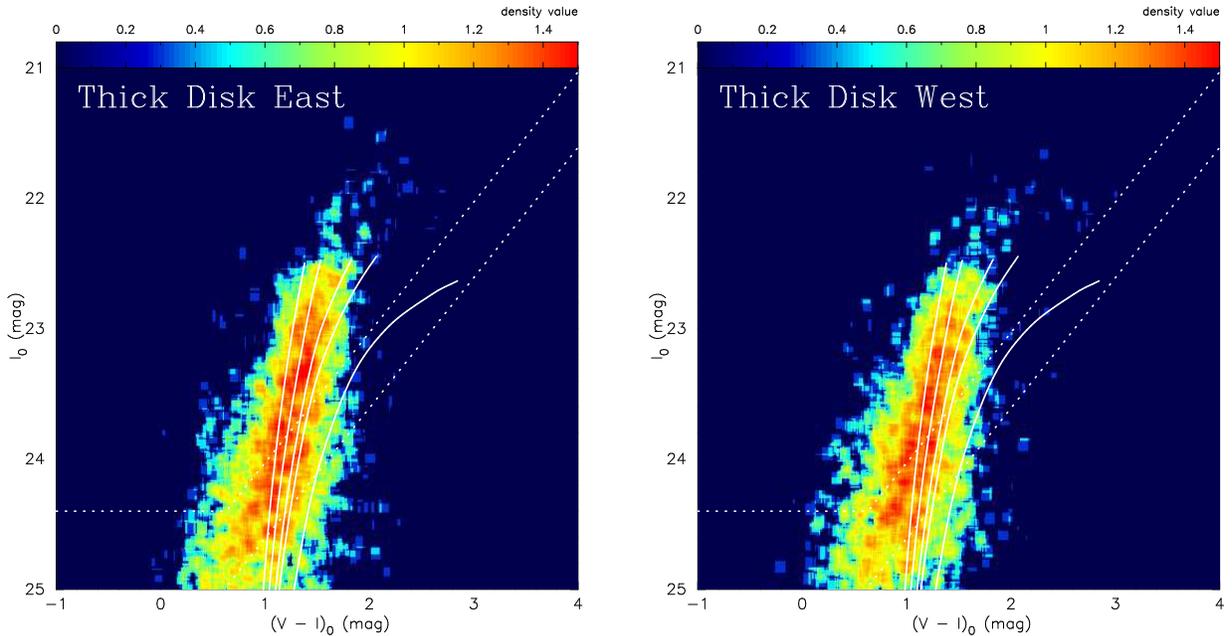}{./f5b.ps}
  \caption[a]{[$(V-I)_0$, $I_0$] color-magnitude diagrams for
  stellar-like sources in TDE (left) and TDW (right) regions of
  NGC~55. The solid lines are theoretical isochrones
  \citep{VandenBerg06} of age 12~Gyr and  [$\alpha$/Fe]$=+0.3$ spanning
  the metallicity range [Fe/H]$ = -2.31$, $-1.71$, $-1.31$, $-1.14$ and
  $-0.71$. The dotted lines denote the full ranges of the 50\% and 80\%
  completeness levels.
  } 
  \label{fig:cmd_inner}
 \end{center}
\end{figure*}

Next, we compare stellar populations between TDE and TDW,
based on their CMDs shown in Figure~\ref{fig:cmd_inner}. There are
well-populated RGB stars in both CMDs, suggesting that a dominant
population in both structures is old and metal-poor
with [Fe/H]$ < -1$. Furthermore, it is remarkable that AGB stars
exist above the TRGB magnitude of $I_0 = 22.48$ as also shown in the previous
section. The AGB population of TDE seems to be more sharply rising at
$(V-I) \sim 1.6$ compared to that of TDW. If this feature is real, it
indicates that the stellar population of TDE is different from that of
TDW: the former is possibly several Gyr younger than the latter.
However, color distributions of both TDE and TDW at $21.00 <
I_0 < 22.48$ have a similar shape and peak at $(V-I)_0 \sim 1.7$.
Also, the ratio of the number of AGB stars at $21.0 < I_0 <
22.48$ and $0.6 < (V-I)_0 < 2.5$ to that of bright RGB stars at $22.48 <
I_0 < 23.0$ and $0.6 < (V-I)_0 < 2.0$ is evaluated as about 30\% for
both TDE and TDW, where the effect of contaminations is limited
within their Poisson noise. Thus, population difference between
TDE and TDW would not be significant from the current CMD analysis,
so that that bulk populations in TDE and TDW may originate
from the same, pre-existing disk component.

\subsection{Outer Structure}\label{sec:outer}

\begin{figure*}[htpd]
 \begin{center}
  \epsscale{1}
  \plottwo{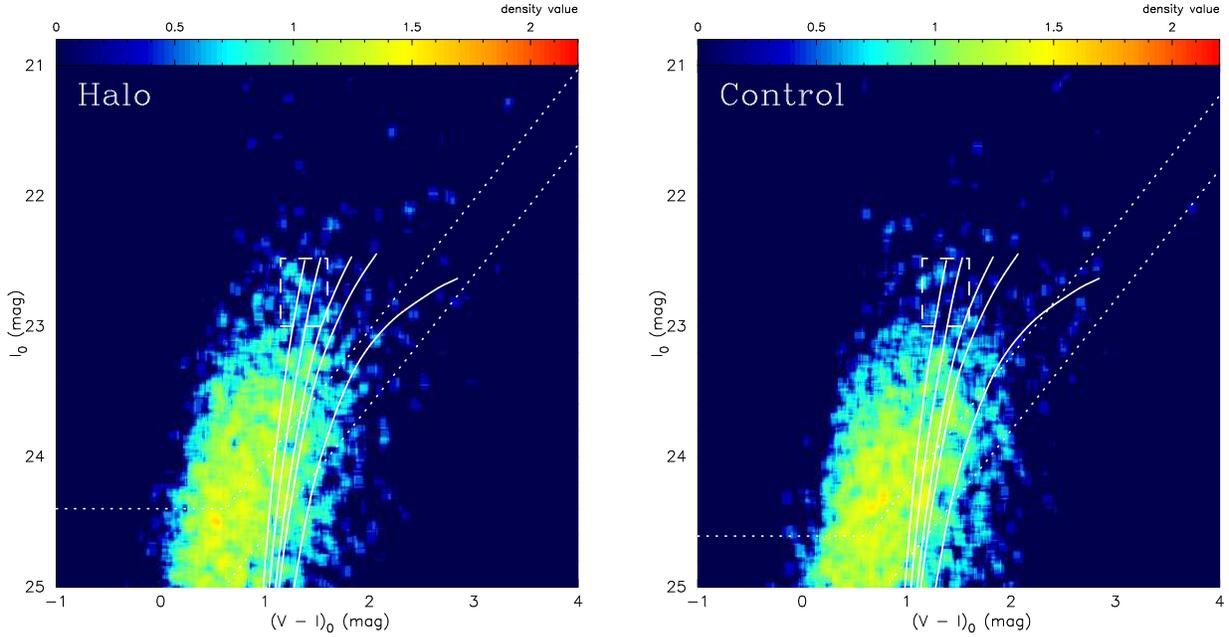}{./f6b.ps}
  \caption[a]{[$(V-I)_0$, $I_0$] CMDs for stellar-like sources in the
  NGC~55's halo (left) and the control field (right). The solid lines
  are theoretical isochrones \citep{VandenBerg06} of age 12~Gyr and
  [$\alpha$/Fe]$=+0.3$ spanning the metallicity range [Fe/H]$ = -2.31$,
  $-1.71$, $-1.31$, $-1.14$ and $-0.71$. The dotted lines denote the
  full ranges of the 50\% and 80\% completeness levels. The existence 
  of metal-poor population in the halo field can be identified
  by comparing the number of RGB stars within the dashed boxes between
  these panels (see text for more details).}
  \label{fig:cmd_halo}
 \end{center}
\end{figure*}

Despite metallicity cuts, the stellar density rapidly declines to a background
level beyond $z\sim5$~kpc, according to the stellar density maps of 
Figure~\ref{fig:map_rgb}. However, metal-poor stars are diffusely
distributed in the outer part of the halo.
The left panel of Figure~\ref{fig:cmd_halo} shows a
CMD in the outer part of NGC~55 which is enclosed by the largest
rectangle in the metal-poor stellar density map of
Figure~\ref{fig:map_rgb}, whereas the right panel presents a
CMD in the control field with the same area of a field-of-view as the
left panel. Although both of the CMDs contain numerous background
contaminations below $I_0 \sim 23$, the CMD of NGC~55's outer part seems
to have a metal-poor population with [Fe/H]$ \lsim -2$ like the diffuse
outer halo of M31 and the Milky Way. The existence of this metal-poor
population in the outer halo can be verified 
by comparing both CMDs at bright and metal-poor RGB part
of $(V-I)_0 \sim 1.2$ and $I_0 \sim 22.8$ (corresponding to dashed
boxes in the CMDs). In fact, the number of the stellar objects inside
this box in the outer halo field of NGC~55 ($n_{\rm halo} = 69$) 
is more than those in the control field ($n_{\rm control} = 43$),
i.e., well beyond the Poisson noise.
This fact that the diffuse halo extends to the outer
part of NGC~55 is consistent with the suggestion of the previous HST
study \citep{Tikhonov05}. However, since its signal-to-noise ratio
is not high (of $\sim 3$), further observations at other wavelengths
are needed to reduce the contamination of unresolved background galaxies
and to confirm the presence of this halo component.

\begin{figure}[htpd]
 \begin{center}
  \epsscale{1}
  \plotone{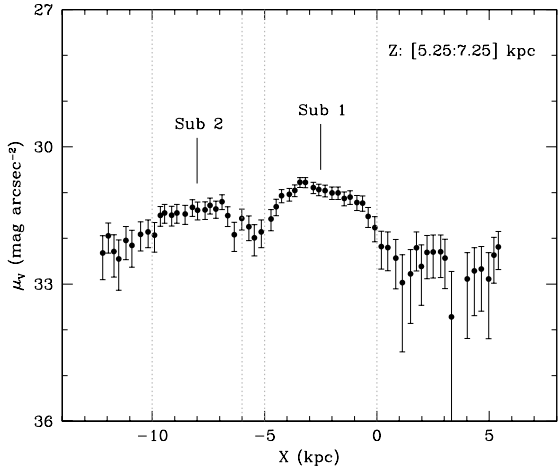}
  \caption[a]{The surface brightness profile in the same direction as the
  major axis at $5.25<z<7.25$~kpc. Two overdense regions bounded by
  vertical gray dotted lines are regarded as substructures,
  Sub~1 and Sub~2, which are also denoted in the stellar density map
  (right panel in Figure~\ref{fig:map_rgb}). 
  }
  \label{fig:sbx}
 \end{center}
\end{figure}

\begin{figure*}[htpd]
 \begin{center}
  \epsscale{1}
  \plottwo{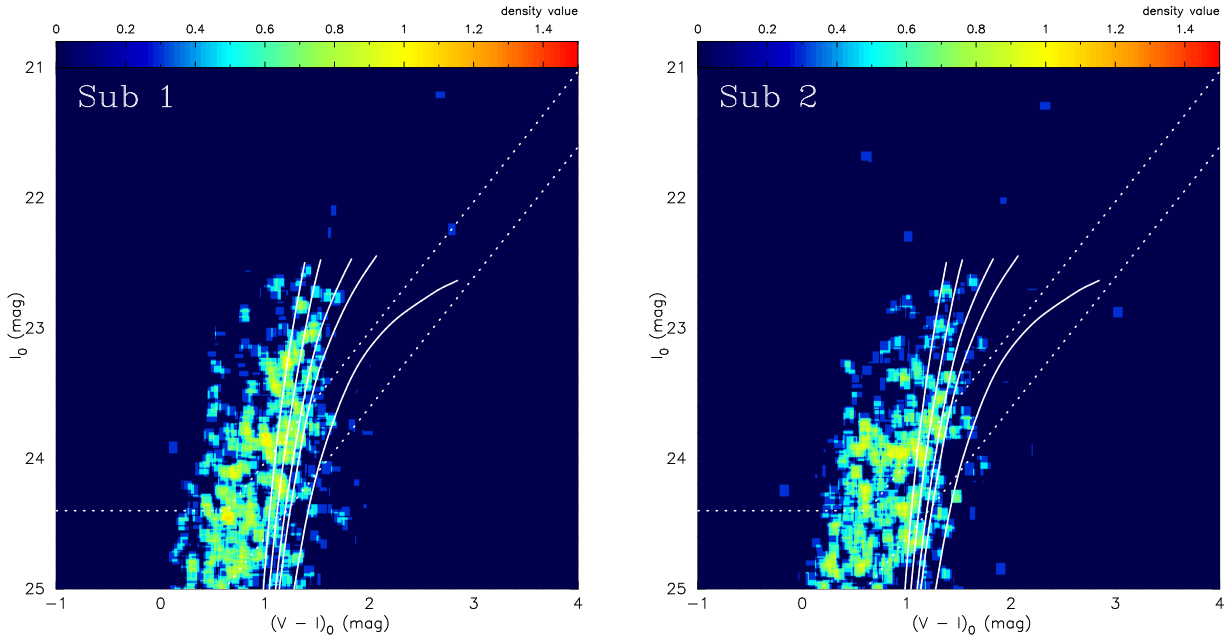}{./f8b.ps}
  \caption[a]{[$(V-I)_0$, $I_0$] CMDs for stellar-like sources in the two
  substructure regions, Sub~1 (left) and Sub~2 (right), of NGC~55's
  halo. The solid lines are theoretical isochrones \citep{VandenBerg06}
  of age 12~Gyr and [$\alpha$/Fe]$=+0.3$ spanning the metallicity range
  [Fe/H]$ = -2.31$, $-1.71$, $-1.31$, $-1.14$ and $-0.71$. The dotted
  linse denote the full ranges of the 50\% and 80\% completeness levels. 
  }
  \label{fig:cmd_outer}
 \end{center}
\end{figure*}

In addition, two regions showing significant excess of stellar density
beyond the level of the diffuse halo can be identified
in the north-east field of NGC~55 (right panel of Figure~\ref{fig:map_rgb}),
which we refer to as Substructure~1 and 2 hereafter.
These halo substructures are more clearly presented in Figure~\ref{fig:sbx},
which shows a background-subtracted surface brightness
profile in the same direction as the major axis at $5.25<z<7.25$~kpc:
the surface brightness is
estimated using the resolved RGB stars selected as $(V_0,I_0) < (25.03,
24.40)$, i.e., brighter magnitude than 80\% completeness limit (see also
Section~\ref{sec:dis}). 
The error bars are estimated from the square root of the number counts 
including both stellar and background flux. 
The profile shows the presence of two
overdense regions beyond the Poisson noise at $x \sim 2.5$~kpc and $x \sim
8$~kpc bounded by vertical gray dotted lines. The CMDs of both substructures
(Figure~\ref{fig:cmd_outer}) show a similar distribution of RGB stars
including the presence of metal-poor stars with $-2.3 \lsim $[Fe/H]$ \lsim -1.7$,
although these substructures are spatially separated as shown in Figure~\ref{fig:sbx}.
The difference of stellar population between these substructures
will be more quantitatively examined based on the analysis of
their MDs in Section~\ref{sec:metal}.

\subsection{Surface Brightness Profile}\label{sec:sbp}

\begin{figure}[htpd]
 \begin{center}
  \epsscale{1}
  \plotone{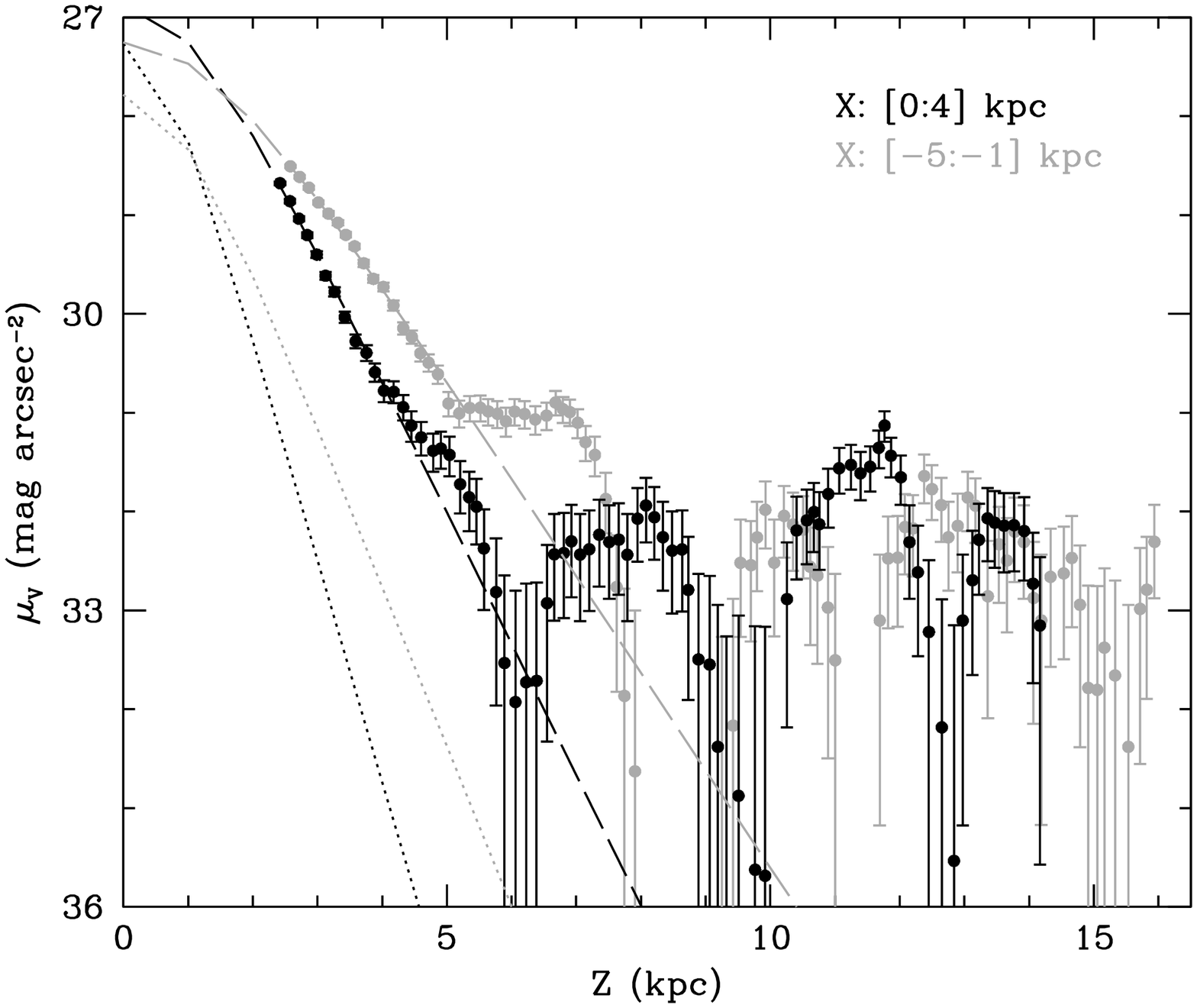}
  \caption[a]{The surface brightness profile in the perpendicular
  direction to the galactic plane at a restricted $x$ range, for the north-western (black) 
  and north-eastern (gray) region of NGC~55. The long-dashed lines denote
  the model distributions of old RGB stars in the form of a locally isothermal
  disk model with a scale height of
  $z_{\rm 0,RGB}^{\rm W} = 1635\pm30$~pc (black) and
  $z_{\rm 0,RGB}^{\rm E} = 2203\pm25$~pc (gray), respectively.
  The dotted lines denote those of AGB stars with a scale height of
  $z_{\rm 0,AGB}^{\rm W} = 969\pm62$~pc (black) and
  $z_{\rm 0,AGB}^{\rm E} = 1341\pm60$~pc (gray), respectively.
  %In addition, the short-dashed line is a power-law
  %very roughly fit to the data of the eastern halo. 
  The background levels for RGB and AGB stars
  are $\mu_{V, \rm back}^{\rm RGB} \sim 30.8 \pm 0.1$ mag arcsec$^{-2}$ and 
  $\mu_{V, \rm back}^{\rm AGB} \sim 33.6 \pm 0.1$ mag arcsec$^{-2}$, respectively. 
  }
  \label{fig:sbz}
 \end{center}
\end{figure}

To further investigate NGC~55's global structure, 
we investigate the distribution of surface brightness
for the resolved RGB stars using those brighter than 80\% completeness limit.
Figure~\ref{fig:sbz} shows surface brightness profiles
converting the summed-up flux counts of selected stars to the surface
brightness in mag arcsec$^{-2}$, and subtracting the surface brightness
of the control field, for which remaining foreground and background
contaminations are removed based on the statistical method
\citep{Tanaka10}. 
The error bars are estimated from the square root of the number counts 
including both stellar and background flux. 
Black and gray circles show the surface brightness profiles
at $0 < x < 4$~kpc (where stars in TDW dominate) and
$-5 < x < -1$~kpc (where those in TDE dominate), respectively.
It is noted that in the latter profile,
there is a prominent overdense structure (corresponding to Sub~1)
at $z \sim 6$~kpc, as also mentioned in the previous subsection. 

It also follows from the figure that the thick disk component is extended
at least up to about 5~kpc, as represented by
isothermal disk models (dashed lines): the vertical
profile of a disk structure is defined as \citep{KS81}:
\begin{equation}
 \Sigma(z) \varpropto sech^2\left(\frac{z}{z_0}\right)
\end{equation}
where $\Sigma(z)$ is the surface brightness or density at a position $z$
above the midplane and $z_0$ is the scale height. As discussed in 
Section~\ref{sec:inner}, the scale height of TDW is $z_{\rm 0,RGB}^{\rm W} = 1635\pm30$~pc 
which is a typical value for a late-type galaxy with $V_{\rm circ} = 110$~km~s$^{-1}$ 
\citep{Yoachim06}. However, on the assumption that NGC~55 has slower rotational velocity 
like $60 \lsim V_{\rm circ} \lsim 90$~km~s$^{-1}$ \citep[e.g.,][]{Puche91}, it has 
a somewhat thicker disk. Nonetheless, it is notable that the
thick disk profile in the eastern part (TDE) has a significantly
larger scale height, $z_{\rm 0,RGB}^{\rm E} = 2203\pm25$~pc, 
than in the western part  (TDW), implying
that the eastern thick disk has been heated up by the interaction with an
accreted dwarf galaxy in the Sculptor group. 

Regarding the surface brightness profile of the bright AGB stars
(fitted with dotted lines in the figure), we obtain the scale heights
of $z_{\rm 0,AGB}^{\rm W} = 969\pm62$~pc and $z_{\rm 0,AGB}^{\rm E} = 1341\pm60$~pc,
respectively, for the north-western and north-eastern regions of NGC~55.
As reported by previous studies of NGC~55 and other galaxies 
\citep{Seth05b,Tikhonov05}, the intermediate-age AGB stars in spiral galaxies show
steeper spatial gradients in their number density than the old RGB stars
and thus are almost absent in the outskirts of galaxies. 
This result strongly suggests the presence of an older component with a larger scale height in 
the outer part of the disk as investigated in \citet{Seth05b}.
Furthermore, the old RGB populations have a scale height similar to typical 
thick disk components, whereas the intermediate-age stellar population has a scale height 
very similar to typical thin disk components; this is consistent with the more systematic 
consideration of \citet{Yoachim06}. Therefore, assuming our AGB population more clearly
traces the thin disk of NGC~55, the eastern part of the thin disk in NGC~55 may have been
dynamically heated by merging satellites, forming a thicker disk as seen
in the eastern part. In addition, 
the fact that the scale height we calculated based on AGB is somewhat larger than 
that of the more inner eastern disk estimated by \citet{Seth05b} supports that 
the outer intermediate-age disk is more significantly expanded. Therefore, NGC~55's 
disk may be extended toward a large height at around 4~kpc from the galactic plane. 

Finally, we investigate the remote outer region of the diffuse halo of NGC~55 at $z \gsim 8$~kpc. 
\citet{Tikhonov05} reported that there is a transitional point from the disk to the halo at 6.5~kpc
based on the deviation of their estimated density distribution from the starlight 
(see their Figure~12). However, considering our surface brightness profile shown in Figure~\ref{fig:sbx} 
and \ref{fig:sbz}, it is likely that the deviation is attributable to the overdense substructure 
of Sub~2, taking into account their HST/WFPC2 pointings on the minor axis of NGC~55. 
In fact, as discussed in Section~\ref{sec:outer}, the outer halo of NGC~55 is substantially diffuse, 
and seems to have a nearly flat profile below the background level, 
$\mu_{V, \rm back}^{\rm RGB} \sim 30.8 \pm 0.1$ mag arcsec$^{-2}$. 
Taking into account this faintness and spatial profile of the NGC~55's halo, its total mass
may be smaller than that of M31, like halos in other late-type galaxies such as M33 and LMC. 
It is noted that the presence of more substructures at a more remote field from the galaxy center
($z\sim12$~kpc) may be likely if the halo of NGC~55 also originates from accretion
of many dwarf galaxies, as suggested for bright spirals like M31 and the Milky Way
\citep{BJ05}. However, to assess this 
picture, further observations of the outer halo of NGC~55 are necessary.

\section{Metallicity Distributions}\label{sec:metal}

\begin{figure*}[htpd]
 \begin{center}
  \epsscale{1}
  \plotone{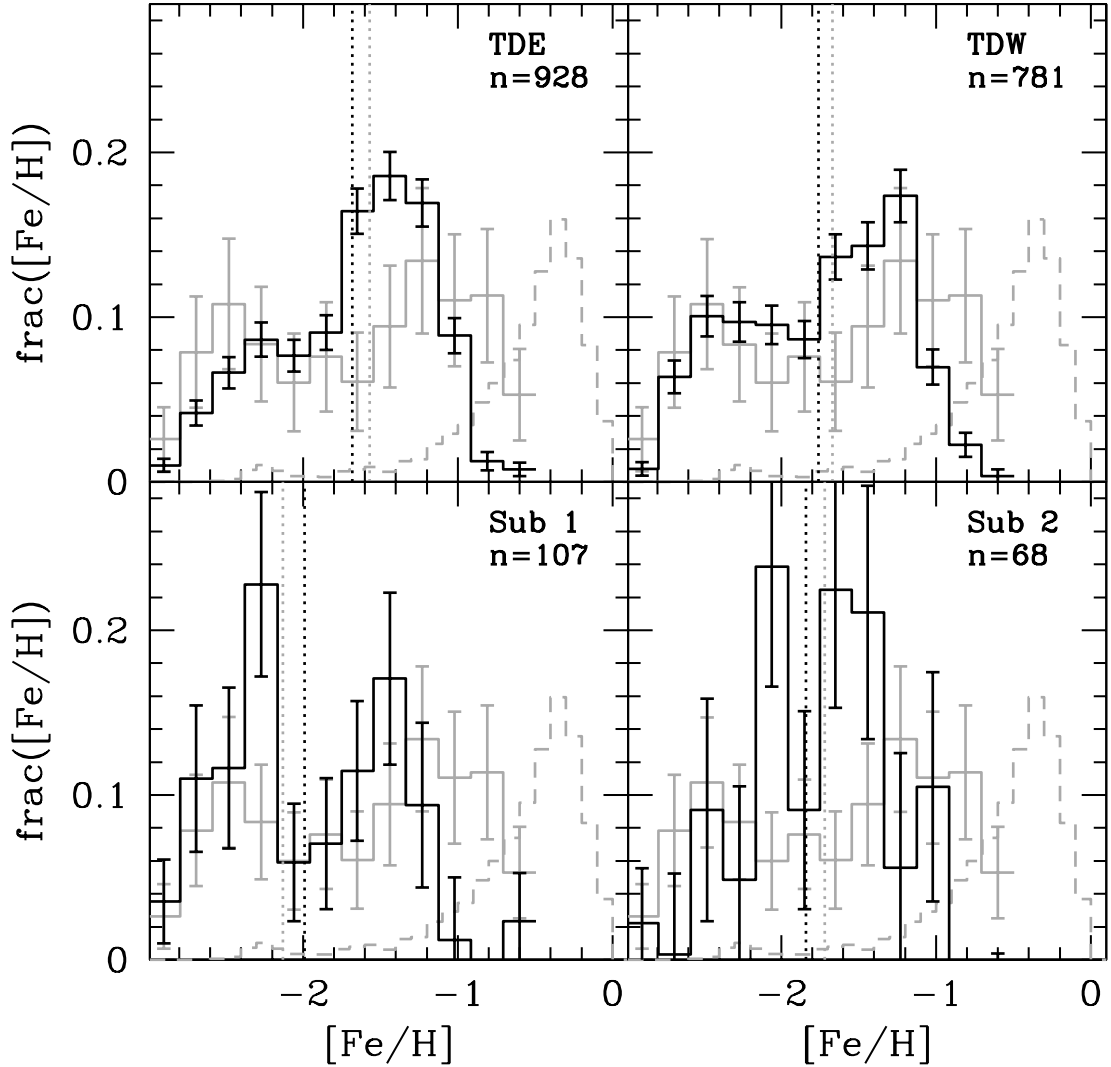}
  \caption[a]{Metallicity distributions for the thick disk (upper panels) and
  substructures (lower panels). The number of stars used to derive these are
  shown in the upper right corner of each panel. Gray histograms correspond 
  to the MD of the control field consisting of 69 objects and
  gray dashed histograms correspond to the MD of M31's Giant Southern Stream
  obtained by \citet{Tanaka10}, which is not sensitive to the [Fe/H]$ > -0.7$ metallicities.
  The vertical dotted lines show the mean metallicity (black) and median 
  metallicity (gray) of each field.
  }
  \label{fig:mdf}
 \end{center}
\end{figure*}

In this section, we discuss the difference of stellar populations
between the above-mentioned substructures, based on the comparison
for the MDs of the RGB stars. To construct MDs, we adopt the Victoria-Regina
theoretical isochrones from \citet{VandenBerg06} \citep[see
also][]{Tanaka10}. In accordance with the interpolation and
extrapolation scheme of \citet{Kalirai06}, we calculate the metallicity 
for each star in the same segment of the CMD, assuming stellar
population with [$\alpha$/Fe]$ = +0.3$, age of 12 Gyr and the same
distance modulus of 26.58 estimated in \S~\ref{sec:dist}. Furthermore,
for secure determinations of MDs, we select the targeted RGB stars
having $22.48 < I_0 < 23.48$ and $1.05 < (V-I)_0 < 2.2$. These selection
criteria allow us to remove a number of contaminations, such as AGB
stars, young stars and background galaxies. After performing the
interpolation procedure for our photometric data, we subtract the
MD of the control field from that of each object field in order to
remove the effects of the remaining contaminations. Since the MD of
the control field shows a nearly flat distribution within the Poisson
errors in the effective metallicity range of $-2.8 \lsim $[Fe/H]$ \lsim -0.7$
(gray histograms in Figure~\ref{fig:mdf}), this subtraction procedure does not
affect the determination for the intrinsic shape of NGC~55's MDs.
In addition, the fraction of the contaminations is estimated as
roughly 30\% in Sub~2 field with the smallest statistic,
whereas it is no more than 6\% in TDE with the highest statistic. 

Figure~\ref{fig:mdf} shows the resultant contaminations-subtracted MDs
for the four substructures we have newly identified in this
study. The vertical error bars denote a nominal uncertainty in each
metallicity bin as derived from the Poisson errors. For reference, we also
plot the MD of the control field (consisting of 69 objects) by a gray
histogram, which clearly shows a nearly flat distribution within the Poisson
errors. The vertical dotted lines correspond to the values of mean (black)
and median (gray) metallicity in each substructure, which are also summarized
in Table~\ref{tab:sub}. It follows that the MDs
of TDE and TDW with high stellar density have a more metal-rich peak
([Fe/H]$_{\rm peak} \sim -1.4$) than those of Sub~1 and 2 with
low stellar density. All the MDs show a somewhat broad
distribution ranging from [Fe/H]$ \sim -3$ to $\sim -1$. For comparison,
we also plot the MD for the Giant Southern Stream observed in M31's halo
\citep{Tanaka10} (gray dashed histogram), which is clearly different
from the MDs in NGC~55. It is also worth comparing with
the MDs of NGC~55's thin disk ([Fe/H]$_{\rm mean}^{\rm thin} \sim -1.0$) 
as obtained by \citet{Seth05b} (see their Figure~11): although
the MDs of TDW and TDE show a similar shape to the MDs of the thin disk,
the former are systematically more metal-poor than the latter, even
taking into account the likely effect of adopting the different
theoretical model of stellar evolution and somewhat large photometric
errors in \citet{Seth05b}. This may suggest some population difference
between our detected thick disk and the inner thin disk in NGC~55.

We also note that the MD in the individual region of NGC~55 shows
a slightly different profile from each other:
the MD of Sub~1 seems to show a higher fraction of metal-poor stars
than that of Sub~2 and the MD of TDE has slightly a smaller fraction of
metal-poor stars with [Fe/H]$ \lsim -2$ than that of TDW.
These differences in MDs may reflect different stellar
population in each substructure. To quantify the difference in these MDs,
we employ a two-sided Kolmogorov-Smirnov (KS) test for the cumulative form
of MDs. The calculated KS probability for
the null hypothesis that all of the four populations originate from the same
stellar population is only less than 1\%: this is a maximum probability
for all of the KS tests. Thus, these MDs are statistically different,
thereby implying that stellar population in each substructure is different.
However, we note that we cannot get rid of the degeneracies between
metallicity and age when using isochrones to infer metallicity.

\begin{deluxetable*}{lcccccc}
 \tablewidth{0pt}
 \tablecaption{Fundamental properties of the substructures\label{tab:sub}}
 \tablehead{
 \colhead{Name} & \colhead{$\mu_V$} & \colhead{[Fe/H]$_{\rm mean}$}
 & \colhead{Standard Div.} & \colhead{[Fe/H]$_{\rm med}$} &
 \colhead{Quartile Div.} & \colhead{Error} \\
  & (mag arcsec$^{-2}$) & (dex) & (dex) & (dex) & (dex) & (dex)
 }
 \startdata
 Thick Disk East & $29.0 \pm 0.03$ & $-1.68$ & $0.54$ & $-1.57$ & $0.37$ & $0.14$ \\
 Thick Disk West & $28.7 \pm 0.02$ & $-1.76$ & $0.58$ & $-1.67$ & $0.46$ & $0.14$ \\
 Substructure~1 & $30.8 \pm 0.11$ & $-1.99$ & $0.59$ & $-2.13$ & $0.43$ & $0.13$ \\
 Substructure~2 & $31.2 \pm 0.17$ & $-1.84$ & $0.42$ & $-1.72$ & $0.30$ & $0.13$
 \enddata
\end{deluxetable*}

\section{Discussion and Concluding Remarks}\label{sec:dis}

Our Suprime-Cam observation of NGC~55 has revealed that this late-type
galaxy holds extended galactic structures well beyond its bright disk
component, namely TDE, TDW, Sub 1, Sub 2, and a diffuse halo.
We discuss here the possible origin of these extended components
in NGC~55.

As already shown in Section 5, the stellar population in each of these
structures seems to be statistically different from each other, as
deduced from the analysis of these MDs. However, taking into account
the spatial distribution of Sub 1 and Sub 2 as well as the difference
of stellar populations between TDE and TDW, there is a possibility
that both Sub 1 and Sub 2 may be associated with some merging event
which also gives rise to disk thickening in its east part.
If this conjecture is the case,
then the MD of TDE should be a combination of the MD of TDW
and that of either Sub 1 or Sub 2, which is however unlikely
as it follows from Figure 10. Thus, the difference in
stellar populations between TDE and TDW may be attributed to
another merging event, which is not related to the origin of
Sub 1 and Sub 2.

Next, we compare the stellar populations of NGC~55's substructures with
those of M31's halo. The MDs of NGC~55's substructures show
a very different shape from those of M31's Giant Southern Stream,
thereby suggesting that stellar population is quite different
between NGC~55 and M31. Furthermore, the MDs of other substructures
in M31 such as Stream~C and D with low stellar density
\citep{Tanaka10} are also different from those of NGC~55.
These results may suggest the difference in the formation processes of
stellar halos among different Hubble types of galaxies. 

\begin{figure}[htpd]
 \begin{center}
  \epsscale{1}
  \plotone{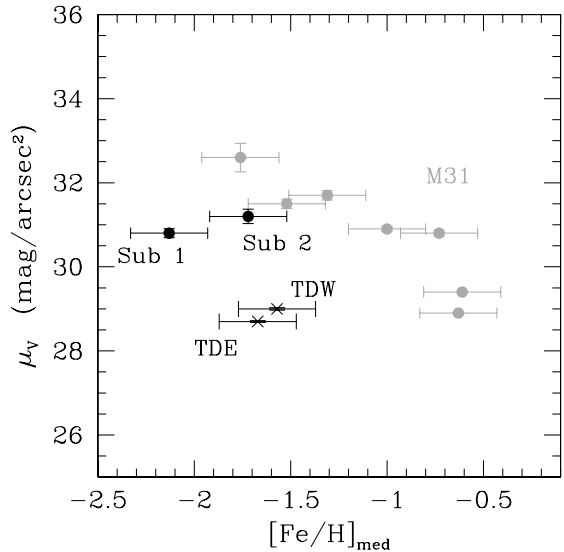}
  \caption[a]{The median metallicity against surface brightness for the 
  substructures of NGC~55 (filled black circles) and M31 (filled gray circles). 
  Black crosses denote the disturbed and undisturbed thick disk structures of NGC~55. 
  Metallicity uncertainties ($\pm 0.2$ dex) are derived from the error analysis
  given in \citet{Tanaka10}. It follows that tidal debris with higher
  surface brightness tend to be more metal-rich in both galaxies and that
  NGC~55's trend is systematically shifted towards a more metal-poor end than
  that for M31's halo.
  }
  \label{fig:femu}
 \end{center}
\end{figure}

Previous studies of M31's stellar halo show a positive correlation between
metal abundance and surface brightness for substructures observed
in M31's halo \citep{Gilbert09,Tanaka10}, i.e., tidal debris with
higher surface brightness tend to be more metal-rich. This correlation
implies that tidal debris with higher surface brightness may originate from
more luminous and thus metal-rich dwarf satellites, and/or more recent
encounters. For NGC~55's substructures, a similar correlation can be
seen as shown in Figure~\ref{fig:femu}, although the trend is
systematically shifted toward a more metal-poor end than that for
M31's halo. Thus, NGC~55's stellar halo may have formed through merging
of relatively metal-poor dwarf galaxies compared to M31's stellar halo.
Provided that the mass-metallicity relation of dwarf galaxies
in the Sculptor Group is the same as that observed in the Local Group
\citet{Cote00}, NGC~55's halo may have originated from less massive
dwarf galaxies than those for the formation of M31's halo, thereby
implying that NGC~55 itself is less massive than M31 at the current epoch.
This seems to be in agreement with the properties of M33 as well: this late-type
galaxy has a less massive halo than M31 and has some faint substructures
with metal-poor and low surface brightness in its halo
\citep{McConnachie10}.

The current study thus suggests that galaxies with different Hubble types
have different properties of stellar halos as characterized by different
metallicity and surface brightness in their substructures.
However stellar halos in earlier Hubble types have yet been
unidentified, simply because there are only few such early-type
galaxies in the local volume so that current telescopes
are able to resolve stars. On the other hand,
a variety of substructures in the form of tidal streams have been
found around more distant galaxies, where it is noteworthy
that morphologies of tidal streams between early and late type galaxies
are quite different \citep[e.g.,][]{Peng02,Martinez10}.
Resolving these substructures into individual stars by, e.g.,
Thirty Meter Telescope and/or Extremely Large Telescope,
would provide further insights into the formation and evolution of
stellar halos as a function of the Hubble sequence.

\medskip

As part of our survey of stellar halos in external galaxies, 
we have observed the north part of the stellar halo in NGC~55, 
using deep and wide-field $V$- and $I$-band images taken with Subaru/Suprime-Cam. 
On the basis of the analysis of the CMDs compared with theoretical isochrones, 
we have obtained the following major results:

\begin{enumerate}
\item
We have found that the stellar populations at $z 
\gsim 5$~kpc above the disk plane are dominated by old RGB stars with
lower metallicity than M31 and the Milky Way. We derive a TRGB-based distance
modulus to the galaxy of $(m-M)_0 = 26.58 \pm 0.11 (d = 2.1 \pm 0.1 {\rm Mpc})$. 

\item
Based on the density map of NGC~55's stellar populations, 
we have found asymmetric thick disk features in the north region of NGC~55, 
which we refer to as TDE for a disturbed thick disk in the east and TDW for an undisturbed 
thick disk in the west.
We have also identified two overdense substructures at $z \sim 6.5$~kpc, 
which we refer to as Substructure~1 and 2 in this work.
These substructures may correspond to remnants of merging events of small galaxies
associated with the formation of NGC~55's halo.

\item
Detailed comparisons between the photometric data for halo regions of NGC~55 and 
that of the control field suggest the presence of a diffuse metal-poor halo 
extended out to at least $z \sim 16$~kpc. 
However, it is yet unclear to what extent this faint component is actually distributed.

\item
The stellar density distribution perpendicular to the galactic plane of 
NGC~55 is described by a locally isothermal disk at $z \lsim 6$~kpc 
and a diffuse metal-poor halo 
with $\mu_{\rm V} \gsim 32$ mag arcsec$^{-2}$ at higher $z$,
where old RGB stars dominate.

\item
From the photometric comparison of RGB stars with the theoretical 
stellar evolutionary model, we have obtained the MDs in the extraplanar regions.
Both TDE and TDW show the peak, average and median metallicity of
[Fe/H]$_{\rm peak} \sim -1.4$, [Fe/H]$_{\rm mean} \sim -1.7$ 
and [Fe/H]$_{\rm med} \sim -1.6$, respectively. 
In contrast, Substructure~1 and 2 show more metal-poor features than the thick disk structures. 
Furthermore, the low KS probabilities indicate that all the MDs are 
statistically different, suggesting that the stellar population in each substructure
may have a different origin.

\end{enumerate}

\acknowledgments

Data reduction and analysis were carried out on general common use computer
system at ADAC (Astronomical Data Analysis Center) of the National
Astronomical Observatory of Japan.
This work has been supported in part by a Grant-in-Aid for
Scientific Research (20340039) of the Ministry of Education, Culture,
Sports, Science and Technology in Japan.

\end{document}